# SHORT-TIME DEEP-LEARNING BASED SOURCE SEPARATION FOR SPEECH ENHANCEMENT IN REVERBERANT ENVIRONMENTS WITH BEAMFORMING

Alejandro Díaz, Diego Pincheira, Rodrigo Mahu, and Nestor Becerra Yoma
Speech Process. and Transm. Lab., Elec. Eng. Dept., U. de Chile, Santiago, Chile.
nbecerra@ing.uchile.cl    http://www.lptv.cl

## ABSTRACT

The source separation-based speech enhancement problem with multiple beamforming in reverberant indoor environments is addressed in this paper. We propose that more generic solutions should cope with time-varying dynamic scenarios with moving microphone array or sources such as those found in voice-based human-robot interaction or smart speaker applications. The effectiveness of ordinary source separation methods based on statistical models such as ICA and NMF depends on the analysis window size and cannot handle reverberation environments. To address these limitations, a short-term source separation method based on a temporal convolutional network in combination with compact bilinear pooling is presented. The proposed scheme is virtually independent of the analysis window size and does not lose effectiveness when the analysis window is shortened to 1.6s, which in turn is very interesting to tackle the source separation problem in time-varying scenarios. Also, improvements in WER as high as 80% were obtained when compared to ICA and NMF with multi-condition reverberant training and testing, and with time-varying SNR experiments to simulate a moving target speech source. Finally, the experiment with the estimation of the clean signal employing the proposed scheme and a clean trained ASR provided a WER 13% lower than the one obtained with the corrupted signal and a multi-condition trained ASR. This surprising result contradicts the widely adopted practice of using multi-condition trained ASR systems and reinforce the use of speech enhancement methods for user profiling in HRI environments.

## 1- INTRODUCTION

The capabilities of robotic and autonomous systems to process and understand spoken language have been growing in the latter years with the use of deep learning and statistical techniques. These techniques can complete numerous tasks, such as speech recognition, speech enhancement, spatial filtering, and speech source separation. Speech source separation is very relevant for human-robot interaction (HRI) in more complex acoustics scenarios (see Fig.1.a) and smart assistants, where the spoken commands to these devices can be of short duration, such as "Alexa" or "Hey Siri" [1], and may take place in the presence of noise sources (see Fig. 1.b). Additionally, the complexity of speech processing is partially defined by the scenario where they occur [2, 3]. For instance, smart speakers are capable of speech enhancement and recognition in a static scenario (Fig. 1.b), where the device lies in a living room, and the speech sources are usually static but could also be moving. A more challenging scenario is shown in Fig.1a where there is a moving robot as well as moving sources. This scenario can be considered a generalization of the smart speaker one and requires more robust techniques that can be able to capture the dynamics of the acoustic environment.

### 1.1 The classical source separation problem

For a multiple channel microphone array, the captured signals can be represented as $Y = [y_0(t), y_1(t), \ldots, y_m(t), \ldots, y_{M-1}(t)]$, where $0 \leq m < M$, $M$ is the total number of microphones and $Y$ is the captured signal matrix. Consider that there are $K$ sources $s_k(t)$, where $0 \leq k < K$. Each microphone $m$ receives the sum of all the $K$ sources, which yields to:

$$y_m(t) = s_{0,m}(t) + s_{1,m}(t) + \cdots + s_{K-1,m}(t) = \sum_{k=0}^{K-1} s_{k,m}(t) \qquad (1)$$

where $s_{k,m}(t)$ denotes source signal $s_k(t)$ received at microphone $m$. This representation can be compressed as a function $y_m(t) = f[s_{k,m}(t)]$ that depends on the position of each source $k$. If we assume that the function is linear, a mixing model can be obtained: $Y = WS$. As defined above, $Y$ is the captured signal matrix of size $M \times 1$ per each time sample $t$, where $0 \leq t < T$ and $T$ is the length in samples of the signal interval being analyzed. Consequently, $Y$ can be interpreted as a matrix of size $M \times T$, $W$ is the $M \times K$ *basis* or *mixing* matrix and $S$ is the $K \times T$ source matrix.

Generally speaking, methods that can exploit the fundamental statistical properties of speech signals (e.g. non-stationary, non-white, and non-Gaussian) have been very popular to tackle speech-related tasks [4]. Independent Component Analysis (ICA), and Non-negative Matrix Factorization (NMF) are typical examples of this kind of approach. These statistical techniques are also known as decomposition methods, and all share the same principle that a given matrix $V$ can be factorized into two other matrices $W$ and $H$. Yet each method has its own constraints: ICA assumes the rows of $H$ to be iid (independent and identically distributed) [5]; and, NMF has the constraint that all matrices have to be non-negative [6].

ICA is employed to address the blind source separation problem by assuming that the sources are temporally iid and non-Gaussian. Under these assumptions, ICA estimates a matrix $B$ called demixing matrix that leads to an estimation of the source signals $\hat{S} = BY$. Given the mixing model discussed above, $Y = WS$ and $B$ corresponds to the inverse of $W$. Similarly, NMF is used to approximate matrix $Y$ by considering that matrices $W$ and $H$ are non-negative [6]. Both methods, ICA and NMF, have been used to tackle multiple speech-related tasks as found in the literature. In [7], the authors used NMF to tackle the speech enhancement task in a complex noise environment in combination with a microphone array. A similar approach for speech separation was used in [8], where the authors employed beamforming to accelerate the convergence of the ICA algorithm under static and low to mild reverberant conditions (RT=150 and 300 ms). The results show a poor performance when mild reverberation was present in comparison to low or zero reverberation. In [9], a multi-stage ICA algorithm is presented using time-domain ICA (TDICA) and frequency-domain ICA (FDICA). It is shown that FDICA performs better than TDICA under reverberant conditions, but the permutation problem appears. The authors showed that decreasing the number of bins of the FFT improved robustness to the permutation problem, but it showed poor performance under reverberation. A combined approach using NMF and K-means clustering for source separation was presented in [10], where the authors tested their method under multiple datasets. The method showed good performance for linear mixtures of speech and music, but poor performance for convolutive mixtures. In [11], a NMF based method was presented for source separation. The authors presented a Bayesian probabilistic framework for multi-source modeling. This framework tackles a significant constraint with the classic NMF methods, which is that the number of sources must be known. In [12], independent vector analysis (IVA), which is a generalization of ICA, improved the performance of source separation in non-stationary scenarios by adding a time-varying parameter in a GMM so that there is no pre-training process. As an important drawback of source separation based on statistical methods, we can mention the dependence on the amount of available data to optimize the objective functions, which in turn imposes restrictions on the applicability to time-varying dynamic environments.

Speech enhancement and source separation are two strongly interrelated problems. As presented in [13], the two terms are interchangeable if we solve the task of extracting one speech source and canceling interferent speakers or noise. Simultaneously, speech enhancement is a more general term that refers to extracting one or more sources in the presence of reverberation, noise, or interferent speakers. Mainly two algorithmic frameworks have been adopted to tackle the speech enhancement and source separation problem, microphone array processing and blind source separation. According to [14], these two converging frameworks share theory and concepts and are used jointly in the literature.

## 1.2 The indoor multiple source problem and reverberation

Reverberation can be modeled as a linear time-invariant system in static acoustic scenarios. Provided a reverberant impulse response (RIR), $h(t)$, the reverberant signal, $x(t)$, can be modeled as $x(t) = \int_0^\infty s(t')h(t-t')\,dt'$, where s$(t)$ is the clean speech. In the frequency domain, the reverberated signal can be expressed as $X(\omega) = H(\omega)S(\omega)$. Reverberation and additive noise reduce the intelligibility of speech and the accuracy of ASR systems, and indoor acoustic scenarios may be highly reverberant and noisy. Classical speech source separation techniques such as ICA and NMF lose effectiveness under reverberation [4, 8, 9, 10]. As expressed above, reverberation generates a convoluted mixing process that is highly time-variant. It also increases the autocorrelation of the original speech signals. Moreover, reverberation remains a challenge for ASR in indoor environments, and many approaches have been proposed to reduce its effect. In [15], Weight Prediction Error (WPE) method was proposed. WPE consists of a statistical speech dereverberation approach that has proven to be very effective and that models the reverberation phenomenon as an autoregressive (AR) model [16, 17, 18, 19]. This technique has widely been employed elsewhere and attempts to dereverberate each frequency bin in the Short-Term Fourier Transform (STFT) of a given signal by using a frequency-dependent linear filter.

Multi-condition training of ASR systems is a widely adopted approach used in reverberant and noisy scenarios. Multi-condition ASR training has shown improvements over clean training [20, 21, 22] with noisy and reverberant signals. Also, it has shown improvements when used in audio-visual speech recognition [23], and spoofing detection [24]. In [20], the authors employed a denoising autoencoder (DAE) speech enhancement stage before ASR. The DAE was trained using reverberant speech as the input and clean speech as the target. The results showed that using only multi-condition training on the ASR outperformed the use of DAE with clean training on the ASR. However, enhancing or cleaning the target speech provides some benefits from the user profiling point of view. To achieve effective collaboration with people, robots need to detect and profile the users with whom they will interact and to modify and adapt their behavior according to the learned models they develop for each user. HRI requires modeling and recognizing human actions and capabilities, to unveil the intentions and goals behind such actions, and to determine the parameters that characterize the social interaction. If robots have a better understanding of each user, they can adapt their behavior concerning users' characteristics and preferences to improve user satisfaction and robot acceptance. At this point, it is worth highlighting that user profiling may be analyzed from a physical, cognitive, and social interaction point of view [25].

Physical profiling - This domain comprises user characteristics that are related to the human body and movements in the space. Consequently, profiling users corresponds to sensing the motion capabilities that are related to the interaction process and the intended movements in the space.

Cognitive profiling-Beyond the classification of observable human physical activities that result from the interaction with the external world, there is the need for predicting, detecting, and recognizing the intents of the observed agent [26]. The capability to infer and recognize the individuals' intentions, desires, beliefs, internal states, personality, and emotions is often referred to as Theory of Mind (ToM) [27].

Social profiling-For successful social interactions with people, robots must recognize and interpret the social cues displayed by a human [28]. Social signals can be defined as observable behaviors that produce behavioral changes during the interaction [29].

User profiling in the physical, cognitive, and social domains is crucial and, in this context, social robots should observe multimodal inputs from human teammates. However, some mode inputs such as physiological signals require wearable sensors that may be invasive from the user´s point of view. Also, image processing may not be always possible depending on the operating conditions. Speech conveys a huge amount of linguistic and paralinguistic (e.g. prosody) information. Beyond voice commands to robots, speech is a window to the psychological, physical, and emotional condition of humans. Nevertheless, speech analysis and processing are very sensitive to noise environments (including the "cocktail party" effect), time-varying acoustic channel, and reverberation in time-varying scenarios. Consequently, effective enhancement or cleaning methods make it possible to profile users by making use of their voices.

**1.3 Deep learning, source separation and speech enhancement**

Convolutional neural networks have been employed to extract representations and features from speech signals [30]. Also, CNN has been used in real-time speech applications [31]. One recent most notable CNN architecture is ResNets [32] that tackles the deep network problem that takes place when stacking too many layers in a neural network resulting in a degradation of the results. ResNet architecture is based on a CNN architecture, i.e. VGG [33] and adds a shortcut connection to a convolutional block. This connection encircles multiple convolutional and dense layers easing the computational cost without losing performance and mitigating the degradation problem [32, 34]. These connections ensure that a new higher layer will perform at least as good as the lower layer. If $Q(x)$ is the function that a section (i.e. one or more layers) of a CNN must approximate to yield good results, then it can asymptotically approximate a residual function $P(x) = Q(x) - x$. If the section of the CNN needs to approximate the identity function, all the weights must be set to zero. The residual connection makes it trivial for the network to do nothing to the input if that is the best thing to do. This architecture has been widely used for multiple tasks in image processing and other fields.

Another notable architecture is the fully convolutional network (FCN) [35]. This architecture uses only convolutional layers, without any dense or fully connected layer. A dense layer is replaced with 1x1 convolutions layers. The advantage of using only convolutional layers is that the input size can be arbitrary since the architecture is only fixed-sized filters independent of the image size. Also, it is expected that the temporal correlation of frames is easily represented in the deeper layers since the filters keep the temporal coherence of frames. Differently, in a dense or fully-connected layer, all the frames are processed and related to all other frames within the analysis window. A recent approach based on FCN is the temporal convolutional network (TCN) [36]. This architecture introduces stacked one-dimensional dilated convolutions that can replace a recurrent network avoiding the use of gating mechanisms. A characteristic of TCN is the low number of parameters needed to obtain comparable results to other recurrent networks [37] such as LSTM.

Source separation and speech enhancement have improved with the use of deep learning [38, 39]. Different deep-learning approaches have been employed such as in [40], where a CNN-LSTM model for multi-speaker source separation was presented using a multi-stage feature extraction with CNN and LSTM autoencoders. The extracted features were concatenated before multiple dense layers. They concluded that the use of both CNN and LSTM autoencoders provided better results than each one individually. Another approach is presented in [41], where the authors described an LSTM architecture that processes spectral and spatial information. The spectrogram was concatenated with spatial and directional features related to the location of the sound sources. With the employment of attention mechanisms, the weights were trained to optimize the use of directional, spatial, and spectral features. In [42], a FCN encoder-decoder network with skip connections is evaluated in data from the REVERB challenge. The FCN encoder-decoder network was used for speech dereverberation and denoising by processing spectrograms as images with two-dimensional convolutional layers. Similarly, TCNs that uses one-dimensional convolutional layers have also been employed for speech enhancement of reverberant signals [43].

## 1.4 Compact bilinear pooling

Deep learning architectures can interpret the combination of multiple signals or information sources. There are several techniques to perform information fusion, such as concatenation, pair-wise multiplication, and bilinear pooling. The performance of bilinear pooling has shown advantages over other fusion techniques, as is studied in [44], where the authors compared different techniques for multimodal fusion in an emotion recognition task. They contrasted bilinear pooling with concatenation, element-wise addition and element-wise multiplication, and bilinear pooling outperformed all of them. Moreover, bilinear pooling combined with deep learning architectures has effectively processed: text and audio [44]; image and audio [45, 46]; text, audio, and video [47]; and multiple audio features [48]. Surprisingly, the application of bilinear pooling has not been exhaustively applied to multiple acoustic sources. An acoustic scene classification scheme that uses bilinear pooling with unimodal information is presented in [48]. The authors extracted harmonic and percussive features from speech spectrograms using a CNN. Bilinear pooling was employed to mix the extracted features and classify acoustic scenes.

Bilinear pooling is as simple as an outer product over two vectors establishing multiplicative relationships between each element of these vectors. The most notorious problem with this technique is that it yields a high dimensionality matrix. Compact bilinear pooling (CBP) tackles that problem with a sampling-based approximation of bilinear pooling [49]. It is based on the Tensor Sketch Projection function $\Psi$, which allows us to map a given vector into a low-dimensional space. If $u$ and $w$ are two vectors with the same dimensionality, it is possible to skip the direct computation of the original outer product $u \otimes w$ by computing the convolution of the feature vectors in a lower dimension space by using $\Psi(u \otimes w, h, s) = \Psi(u, h, s) * \Psi(w, h, s)$, in which $h$ and $s$ are vectors of randomly sampled parameters.

## 1.5 Contribution of this paper

The problem of source separation to enhance a target speech in indoor reverberant environments is addressed by combining a TCN with CBP. The adopted model incorporates the acoustic environment response for multiple beamformers that spatially filter the target speech and noise sources. Bearing in mind the applicability to time-varying dynamic environments, the use of shorter analysis time intervals is explored and prioritized. The final objective is to enable distant ASR in time-varying dynamic scenarios such as those found in HRI with social robots and smart speaker applications. It is worth emphasizing that the effectiveness of ordinary source separation methods based on statistical models such as ICA and NMF depends on the analysis window size and cannot handle reverberation environments. The proposed TCN/CBP scheme is virtually independent of the analysis window size, at least when this is larger than 1.6s, which makes more addressable the source separation problem in time-varying contexts. Moreover, improvements in WER as high as 80% were obtained without and with WPE when compared to ICA and NMF with multi-condition reverberant training and testing, and with time-varying SNR experiments to simulate a moving target speech source. Moreover, the experiment with the estimation of the clean signal utilizing the proposed scheme and the clean trained ASR provided a WER 13% lower than the one obtained with the corrupted signal and the multi-condition trained ASR. This result challenges the widely adopted practice of using multi-condition trained systems and is consistent with the use of enhancement methods that can also be useful for user profiling in HRI environments. To the best of our knowledge, the operating restrictions imposed in this research and the proposed solutions have not been exhaustively explored in the literature.

## 2- SOURCE SEPARATION AND INDOOR REVERBERANT TIME-VARYING SCENARIOS

A microphone array is an arbitrary number of microphones whose outputs can be processed and combined to obtain spatial filtering by generating a beamformed signal. The use of a microphone array to perform beamforming can reduce the effect of noise and reverberation. In the case of reverberation, spatial filtering helps to suppress the non-direct path acoustic signals [50]. In delay-and-sum beamforming, samples $y_m(t)$ from each microphone $m$ are delayed by $\tau_{S_p,m}$ samples and then summed, where $S_p$ is the target source. By doing so, the output beamformed signal $b_{S_p}(t)$ in the discrete-time domain corresponds to:

$$b_{S_p}(t) = \sum_{m=0}^{M-1} y_m(t - \tau_{S_p,m}) \qquad (2)$$

A planar wavefront can be assumed if the distance between the microphone array and the sound source is larger than 5-10 times the length of the array [51]. Consequently, the delay for each microphone is given by:

$$\tau_{S_p,m} = \frac{\Delta_m \cdot \sin \phi_{S_p}}{c} \qquad (3)$$

where $\Delta_m$ is the distance between the microphone $m$ and the reference microphone. $\phi_{S_p}$ is the angle of incidence (AOI) corresponding to the source $S_p$, and, $c$ is the propagation speed of sound in the medium [52]. By replacing $y_m(t - \tau_{S_p,m})$ with (1) and changing the summation order, $b_{S_p}(t)$ can be expressed as:

$$b_{S_p}(t) = \sum_{k=0}^{K-1} \sum_{m=0}^{M-1} s_{k,m}\left(t - \tau_{S_p,m}\right) \qquad (4)$$

Consider that, given a source $k$, all the $s_{k,m}$ has the same energy for $0 \leq m < M$. This is a reasonable assumption if the distance between the microphone array and the sound source is larger than 5-10 times the length of the microphone array, as mentioned above, and if the microphones are omnidirectional. The delay-and-sum beamforming scheme will condition the received source $k$ signal energy with gain that depends on the direction it is pointed to, which in turn is a function of $\tau_{S_p,m}$:

$$b_{S_p}(t) = \sum_{k=0}^{K-1} G_{S_p,k} \cdot s_k(t) \tag{5}$$

where $G_{S_p,k}$ denotes the gain of source $k$ when the beamforming is pointing to the source $S_p$. The time delay due to the propagation from the source until the reference microphone can be omitted if a static scenario is considered. Given Fig. 2 with sources $S_0$ and $S_1$, eq. (5) corresponds to:

$$b_{S_0}(t) = G_{S_0,S_0} * s_0(t) + G_{S_0,S_1} * s_1(t) \tag{6}$$
$$b_{S_1}(t) = G_{S_1,S_0} * s_0(t) + G_{S_1,S_1} * s_1(t) \tag{7}$$

According to eqs. (6) and (7), $b_{S_0}(t)$ and $b_{S_1}(t)$ can be obtained by adding $s_0(t)$ and $s_1(t)$ at different SNRs if $s_1(t)$ is considered noise. In a real indoor environment, the microphone array receives the direct path signal from each source but also the corresponding reflections on the walls, floor, ceiling, and other objects. Consequently, eqs. (6) and (7) can be modified to represent the problem in Fig. 2 more accurately:

$$b_{S_0}(t) = h_{S_0,S_0} * s_0(t) + h_{S_0,S_1} * s_1(t) \tag{8}$$
$$b_{S_1}(t) = h_{S_1,S_0} * s_0(t) + h_{S_1,S_1} * s_1(t) \tag{9}$$

where $h_{S_0,S_0}$ and $h_{S_0,S_1}$ are the RIRs observed by the microphone array at the direction of $S_0$ and $S_1$, respectively, when the beamforming is pointing to $S_0$; $h_{S_1,S_0}$ and $h_{S_1,S_1}$ are the RIRs observed by the microphone array at the direction of $S_0$ and $S_1$, respectively, when the beamforming look direction is targeting $S_1$. Observe that an RIR incorporates the direct path delay, so $h_{S_0,S_0}$, $h_{S_0,S_1}$, $h_{S_1,S_0}$ and $h_{S_1,S_1}$ account for the time delay due to the propagation from the source until the reference microphone. Ordinary source separation methods such as ICA and NMF are not designed to address the problem defined by eqs. (8) and (9). Moreover, a more generic model should represent time-varying dynamic scenarios with moving microphone array or sources. In this case, time-dependent reverberation cannot be modeled by convolving a given RIR with the original signal. However, time-varying scenarios like those in Figs. 1-2 can still be approximated with eqs. (8) and (9) by considering the RIRs as fixed in time intervals where the whole system can be assumed as quasi-static:

$$b_{S_0}(t) \cong h_{S_0,S_0}(t) * s_0(t) + h_{S_0,S_1}(t) * s_1(t) \tag{10}$$
$$b_{S_1}(t) \cong h_{S_1,S_0}(t) * s_0(t) + h_{S_1,S_1}(t) * s_1(t) \tag{11}$$

The shorter the analysis time interval, the more accurate is the quasi-static hypothesis of the environment. Nevertheless, statistical-based methods lose accuracy when the amount of estimation data is reduced and are not good candidates to address the source separation problem as defined in eqs. (10) and (11). To counteract this limitation, we propose a method based on deep learning that is composed of TCN and CBP that could be trained on multiple conditions and accomplish the source separation task in shorter analysis widows. Henceforth $b_{S_0}(t)$ and $b_{S_1}(t)$ will be denoted as $b_0(t)$ and $b_1(t)$, respectively. Observe that $b_0(t)$ and $b_1(t)$ indicate the beamforming signals that point to sources $s_0(t)$ and $s_1(t)$, respectively, where $s_0(t)$ is the target clean speech source and $s_1(t)$ corresponds to the noise source. Accordingly, $h_{S_0,S_0}$, $h_{S_0,S_1}$, $h_{S_1,S_0}$ and $h_{S_1,S_1}$ will be denoted as $h_{00}$, $h_{01}$, $h_{10}$ and $h_{11}$, respectively.

# 3- PROPOSED SOLUTION

Speech enhancement based on speech source separation methods in indoor reverberant and dynamic environments became a challenging problem that cannot be addressed by conventional methods such as ICA and NMF because the classic mixing model is not adequate. Consider the two sources (i.e. target speech and noise) and the corresponding beamforming in Fig.2, and the model expressed in eqs. (10) and (11), $B_0(\omega)$ and $B_1(\omega)$ denote the spectrograms of $b_0(t)$ and $b_1(t)$, respectively. To address this challenge, we propose a deep-learning neural network architecture combined with CBP to obtain a de-reverberated and denoised version of the target speech source signal (see Fig. 3). The proposed solution copes with the restriction that the acoustic scenario can be time-varying and dynamic.

We used one-dimensional fully convolutional blocks with skip connections and increasing dilation rate ($d$) to extract features from the spectrograms (Fig. 3). Each convolutional block is composed of two convolutional layers, which are composed of a set of multiple learnable filters that have a fixed dimension (Fig. 4). These filters are convoluted with the input to produce a new feature map. Typically, in image processing, convolutional layers have 2-dimensional filters and they are applied horizontally and vertically. Spectrograms are temporal-frequency representations of a signal that can be visualized as an image. However, the time and frequency axis are not symmetrical. We used 1-dimensional convolution blocks (called "1-D Conv" in Fig. 3) for our task because reverberation can be modeled on each Short-Time Fourier Transform (STFT) bin trajectory individually [15]. In this context, each spectrogram was interpreted as an array of multiple STFT trajectories or channels. The number of trajectories or channels corresponds to the number of frequency bins, i.e. 257 in our case, and the analysis window or STFT trajectory length is given in terms of the number of frames.

In Fig. 4, the 1-D dilated convolution block is shown in detail. They are composed of two convolution layers with ReLu activation and batch normalization, with one skip connection from the input of the block to the output of the batch normalization. Both convolutional layers have the same dilation rate and filter length. Batch normalization was incorporated after every layer activation to accelerate convergence and prevent overfitting. The batch normalization output is summed to the skip connection resulting in the input of the next block.

The size of the filters determines the receptive field of the input. Increasing the receptive field is desirable to cancel reverberation. However, there is a trade-off between the filter size and the computational load. Since every filter element is a trainable parameter, a larger filter size will increase the computational load and the required amount of training data. A strategy to soften this restriction is to use a dilated convolutional filter [53]. The 1-D convolutional blocks are stacked in the architecture with different dilation rates. In Fig. 3, the first 1-D block consists of 257x257 dimensional filters of length three that move along the time axis for each dilated convolution within the block. The second 1-D block consists of the same number of filters of length three and dilation rate 2; the third 1-D block is the same with dilation rate 4. These 1-D blocks increase evenly in dilation rate until 8, which corresponds to the last 1-D block before CBP.

After processing $B_0(\omega)$ and $B_1(\omega)$ with the top and bottom branches of the proposed architecture in Fig. 3, the CBP block is used to obtain a joint representation of both spectrograms. Then, the resulting features are propagated through the output branch to obtain an estimation of the target speech spectrogram $\hat{S}_0(\omega)$. The dimension of the CBP output is 257 channels x number of frames. The output branch is composed of one 1x1 convolutional layer, a 1-D convolutional block, and a 1x1 convolutional layer. The first 1x1 convolutional block uses ReLu activation, while the 1-D convolutional block and the last 1x1 convolutional layer uses linear activation.

# 4- EXPERIMENTS

The proposed method was evaluated employing experiments with mainly the clean ASR system. The results with the clean trained ASR engine are particularly representative because the presented TCN/CBP scheme delivers an estimation of the original clean speech source. However, comparative results with the multi-condition trained ASR engine were also obtained. The proposed scheme was compared to ICA and NMF according to the dependence on the length of the analysis time interval and to the effectiveness to carry out source separation in the presence of reverberation. The experiments were carried out with four datasets: first, the database that simulates two sources, i.e. speech and restaurant noise, without reverberation; second, one that simulates the speech and noise source in matched reverberant condition, i.e. the TCN/CBP scheme was trained and tested with the same RIR; third, one that employs multi-condition RIR´s, i.e. the TCN/CBP scheme was trained and tested with multiple RIRs; and fourth, with multi-condition RIR´s but time-varying SNR to simulate a moving speech source in testing. Multi-condition RIR´s means that the training RIRs were different from the testing ones. ICA and NMF were applied by making use of ICAmatlab [54] and FASST [55] toolkits, respectively.

## 4.1 Data set generation

We used the clean training and dev clean data from the AURORA-4 corpus, i.e. 7138 and 330 clean utterances, respectively, to generate the training and dev sets for the TCN/CBP system. For testing, we employed the 330 clean utterances from the AURORA-4 database.

### *4.1.1 Beamforming signal generation without reverberation*

To generate $b_0(t)$ according to (6), we added restaurant noise to every clean utterance from the training, dev, and testing database at a random SNR between 0dB and 15dB. The corresponding $b_1(t)$ signal, as defined as in (7), was obtained by adding the same restaurant noise at an SNR equal to 3dB lower than the one used for $b_0(t)$.

### *4.1.2 Beamforming signals generation with reverberation*

We simulated four RIRs per each training, dev, and testing utterance with Pyroomacoustics [56], i.e. $h_{00}$, $h_{01}$, $h_{10}$ and $h_{11}$ as defined in (8) and (9). To generate $b_0(t)$ according to (8), we convoluted each clean utterance and the restaurant noise signal with $h_{00}$ and $h_{01}$, respectively. Then, the resulting convoluted signals were added at a random SNR between 0dB and 15dB. The corresponding $b_1(t)$ utterance as defined in (9) was obtained by convolving the same clean utterance and restaurant noise signal with $h_{10}$ and $h_{11}$, respectively. The resulting convoluted signals were added at an SNR equal to 3dB lower than the one used for $b_0(t)$. In the experiments with matched reverberant condition, i.e. RIR-Matched, all the utterances from the training, dev and testing sets employed the same set of $h_{00}$, $h_{01}$, $h_{10}$ and $h_{11}$. In the reverberant multicondition experiments, i.e. RIR-Multicondition, a different set of $h_{00}$, $h_{01}$, $h_{10}$ and $h_{11}$ were generated for each clean utterance at the training, dev and testing dataset. To generate the RIRs, we employed the scheme presented in Fig. 5 in a simulated room of 2.5m x 6.0m x 6.0m (HxWxD) with a uniform distribution variation with a range of ±20% in all the dimensions. The microphone array and sources were positioned at a random location within the room with a variation in the angle of ±30° as shown in Fig 5. The speaker-to-microphone distance was obtained from a uniform distribution between 1.6m and 2.4m. The speaker and microphone were positioned at a random location with the constraint that they were at least at 1m from any wall, floor, and ceiling. A linear four-microphone array was adopted. The microphones were positioned at -11.3cm, 3.6cm, 7.6cm, and 11.3cm with respect to the center array. These dimensions emulated the Kinect microphone array [57]. Each microphone response was modeled as omnidirectional. Delay-and-sum beamforming was simulated with known time-delays as defined in (2). In this condition, $h_{00}$ and $h_{01}$ were obtained by pointing the beamforming to the target source $S_0$, and $h_{10}$ and $h_{11}$ were generated by pointing the beamforming to source $S_1$. The four RIRs for the matched reverberant condition were generated with the mean value of each distribution employed in the simulations.

### *4.1.3 Time-varying beamforming signal generation*

Based on the same scheme presented in Fig. 5, a database was generated to simulate the SNR variation in the case of moving speech sources. A linear movement between 1.0m and 3.0m, and vice-versa, at an average speed of 5.0 km/h with respect to the microphone array was considered. The signal energy was modified at each sample with gain proportional to $1/x^2$, where x is the distance between the speech source and the microphone array. The gain was applied such that the SNR at 2.0m corresponds to the one mentioned at section 4.1.1 for $b_0(t)$. This database is called Time-Varying-SNR.

## 4.2 TCN/CBP training

The proposed deep learning-based solution was trained with the Keras API for TensorFlow [58]. We used Adam optimizer [59] to train our models. The MSE was adopted as a loss function. The number of epochs was high enough to show overfitting and only the best model according to validation loss was adopted for testing. The hyperparameters were tuned with clean trained ASR experiments and the testing data corresponding to $b_0(t)$ without reverberation. The training, dev and testing utterances were segmented in analysis windows. Each analysis window is presented to the deep learning scheme as an independent unit and is composed of a given number of frames: 160, 320, or 640 frames. Each frame corresponds to 25ms with an overlap of 15ms. If the end of the last analysis window does not coincide with the end of a given utterance, then reflect padding was applied [60]. To mitigate the effect of the random initial parameter variation, the TCN/CBP proposed structure was trained three times for each reported experiment. A desktop PC with an Intel i7-7700 processor, 32GB RAM, and a GeForce GTX1650 6GB GPU was employed for training and testing.

## 4.3 ASR system

We constructed two DNN-HMM ASR systems using the tri2b Kaldi recipe for the database. The first system was trained with the original clean data from the AURORA-4 database. The second one was trained with $b_0(t)$, where the multi-condition RIR´s were generated as in section 4.1.2. Except for the training data, the training procedure employed in both systems is the same. First, a GMM-HMM was built by training a monophone system; then, the alignments from that system were employed to generate an initial triphone system; finally, the triphone alignments were employed to train the final triphone system. This recipe also included Mel-frequency cepstral coefficients (MFCCs), linear discriminant analysis (LDA), and maximum likelihood linear transforms (MLLTs). Once the GMM-HMM system was trained, the GMM was replaced by a DNN. The DNN was composed of seven hidden layers and 2048 units per layer each, and the input considered a context window of 11 frames. The number of units of the output DNN layer was equal to the number of Gaussians in the corresponding GMM-HMM system. The reference for the DNN training was the alignment obtained with the GMM-HMM system trained with the clean data and run on the same data. This leads to a better reference for the DNN than using noisy or corrupted speech data [61, 62]. The feature vector consisted of 40 Mel filter bank (MelFB) features, and delta and delta-delta dynamic features, using an 11-frame context window. The DNN was trained initially using the Cross-Entropy criterion. Then, the final system was obtained by re-training the DNN using sMBR discriminative training [63]. For decoding, the standard 5K lexicon and trigram language model from WSJ were used [64]. As a result, the language model was tuned to the task, i.e. it is task-dependent. The WER with the clean testing data gave 2.19% which is competitive with those published elsewhere. The multi-condition ASR provided a WER equal to 7.75% on the testing data corresponding to $b_0(t)$ with multi-condition RIR´s. A desktop PC with an Intel i7-4790 processor, 32GB RAM, and a GeForce GTX980 4 GB GPU was employed for ASR training and decoding.

## 5. RESULTS AND DISCUSSION

Each WER reported here corresponds to an average WER obtained with the three TCN/CBP neural networks that were trained for each case to reduce the random initial parameter variation effect as discussed in section 4.2. According to Table 1, the estimation of the original clean signal, $\hat{s}(t)$ or $\hat{S}(\omega)$, in source separation experiments without reverberation by making use of ICA, NMF and the proposed TCN/CBP scheme led to reductions of 77%, 81%, and 78%, respectively, when compared with the baseline result with $b_0(t)$. NMF and TCN/CBP provided a WER that is just 0.8% and 1.33% absolute higher than the WER obtained with the clean signals, i.e. 2.19% (see section 4.3), respectively. The results of experiments with different analysis window sizes are presented in Fig. 6. Source separation was carried out in each analysis window for ICA, NMF, and the proposed TCN/CBP scheme. As can be seen in Fig. 6, the WER with ICA and NMF was increased by 81% and 34% relative, respectively, when the analysis window was reduced from 640 to 160 frames. Moreover, there is also degradation in WER when ICA and NMF are applied in 640-frame windows instead of the whole utterance. This increase in WER is equal to 21% and 69% with ICA and NMF, respectively. In contrast, TCN/CBP provided a WER that does not degrade when the analysis window is shortened. However, TCP/CBP provides a slightly higher WER when the analysis window size increases. This must be due to the lower number of the resulting training units or the padding effect that becomes more severe when the analysis window size increases. ICA and NMF are based on statistical methods and are highly dependent on the available testing data. In contrast, the parameters of the proposed TCN/CBP are trained with the information provided by the two sources added at two different SNRs and does not require to estimate any further parameters in the source separation testing process.

Figures 7 and 8 show the source separation results in the presence of reverberation conditions as defined in (8) and (9). In Fig. 7, we tested matched reverberation in the training and testing of the proposed deep learning scheme, i.e. with RIR-matched. As can be seen in Fig. 7, ICA and NMF provided even worse results than the baseline system, i.e. with $b_0(t)$. In contrast, the proposed TCN/CBP provided reductions in WER as high as 91% and 88% without and with WPE, respectively. Similar results can be observed in Fig. 8 that presents the WERs with multiple RIRs in the TCN/CBP training and testing, i.e. RIR-Multicondition. It is worth highlighting that the RIRs employed in the testing data were not used in the training procedure. As can be seen in Fig 8, ICA and NMF kept on giving worse results than the baseline system with $b_0(t)$ and TCN/CBP provided reductions in WER as high as 87% and 82% without and with WPE, respectively. Observe that the improvement due to WPE is quite low when compared to the whole reduction in WER due to TCN/CBP. Although not shown here, similar results were achieved with the Time-Varying-SNR dataset. Also, according to Table 2, the performance of TCN/CBP is very independent of the analysis window size, except small variations that may be caused by the ratio between the number of trained parameters and the amount of data units, and the relative padding effect. Observe that when the analysis window size is reduced, the number of training data units increases while the number of free trainable parameters is kept constant, and the percentage of segments with padding is also reduced. At this point, it is interesting to emphasize that the independence of TCN/CBP with respect to the analysis window length makes the proposed approach a candidate to address more complex time-varying dynamic scenarios modeled with (10) and (11).

Figure 9 depicts the spectrograms of the reference clean signal $s_0(t)$, beamforming signals $b_0(t)$ and $b_1(t)$, and the estimated clean speech, $\hat{s}(t)$. As can be seen in Fig 9, TCN/CBP effectively separates the target clean speech source from the noise one and removes reverberation. According to Table 3, CBP leads to a reduction in WER of 1.6% relative when compared with the simple feature concatenation of the TCN outputs in Fig. 3 when the analysis window size is equal to 160 frames. This result suggests that the TCN component in Fig. 3 accounts for most of the observed improvement in recognition accuracy. Moreover, Table 3 also shows that the whole TCN/CBP scheme takes advantage of both inputs $b_0(t)$ and $b_1(t)$. When only one beamforming signal is present worse results are obtained, i.e. an average increase of 6% relative in WER is observed when recognition is performed with $b_0(t)$ only. Observe that $b_0(t)$ was generated with an SNR higher than $b_1(t)$. According to Table 4, similar results were achieved with the Time-Varying-SNR database except for the fact that in average the WERs in Table 4 are 9% higher than in Table 3. This degradation is relatively low if we consider that the Time-Varying-SNR database is a more complex task than the RIR-Multicondition one, and this result confirms basically the pertinence of the proposed short-term source separation approach.

Table 5 depicts a comparison between clean and multi-condition trained ASR systems with the RIR-Multicondition data. Very surprisingly, the experiment with the estimation of the clean signal utilizing the proposed TCN/CBP scheme, $\hat{s}(t)$, and the clean trained ASR provided a WER 13% lower than the one obtained with $b_0(t)$ and the multi-condition trained ASR. This result contradicts the widely adopted practice of using multi-condition trained ASR systems and supports the use of enhancement methods that can also be useful for user profiling in HRI environments. Finally, Table 6 shows the advantages of TCN/CBP when compared with ICA and NMF with respect to training time, testing time, dependence on analysis window size, and robustness to reverberation between.

## 6- CONCLUSIONS

In this paper, we model the source separation-based speech enhancement problem with multiple beamforming in reverberant indoor environments. We observe that more generic models should represent time-varying dynamic scenarios with moving microphone array or sources. Typical examples could be found in voice-based

HRI with social robots or in smart speaker applications. We also highlight that the effectiveness of ordinary source separation methods based on statistical models such as ICA and NMF depends on the analysis window size and cannot handle reverberation environments. To address these limitations, we propose a method based on a temporal convolutional network in combination with compact bilinear pooling. The proposed scheme is virtually independent of the analysis window size at least when this is larger than 1.6s, which is very interesting to address the source separation problem in time-varying scenarios. Also, improvements in WER as high as 80% were obtained without and with WPE when compared to ICA and NMF with multi-condition reverberant training and testing. Similar results were also achieved with time-varying SNR experiments to simulate a moving target speech source. Finally, and very surprisingly, the experiment with the estimation of the clean signal employing the proposed scheme and the clean trained ASR provided a WER 13% lower than the one obtained with the corrupted signal and the multi-condition trained ASR. This result challenges the widely adopted practice of using multi-condition trained systems and strengthen the use of enhancement methods that can also be useful for user profiling in HRI environments.

## ACKNOWLEDGMENTS

This study was funded by grants Conicyt-Fondecyt 1151306 and ONRG N°62909-17-1-2002.

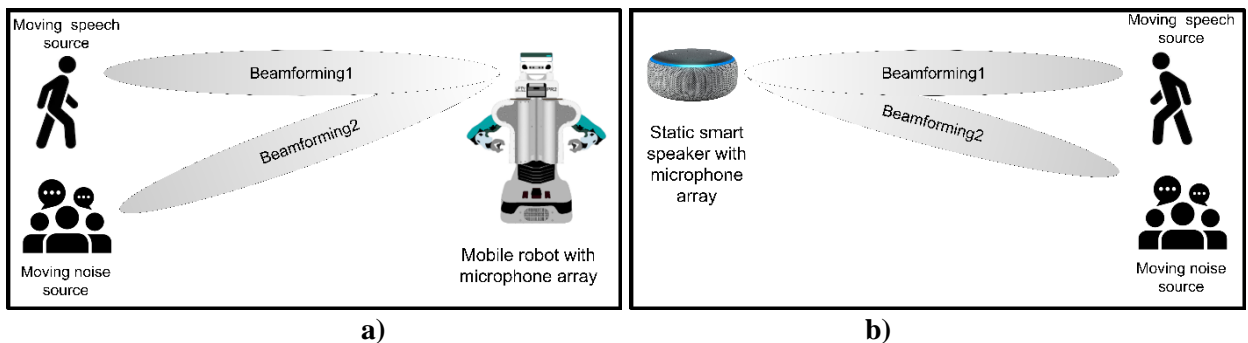

**Figure 1:** Speech and noise moving sources in HRI (a) and smart speaker applications (b)).

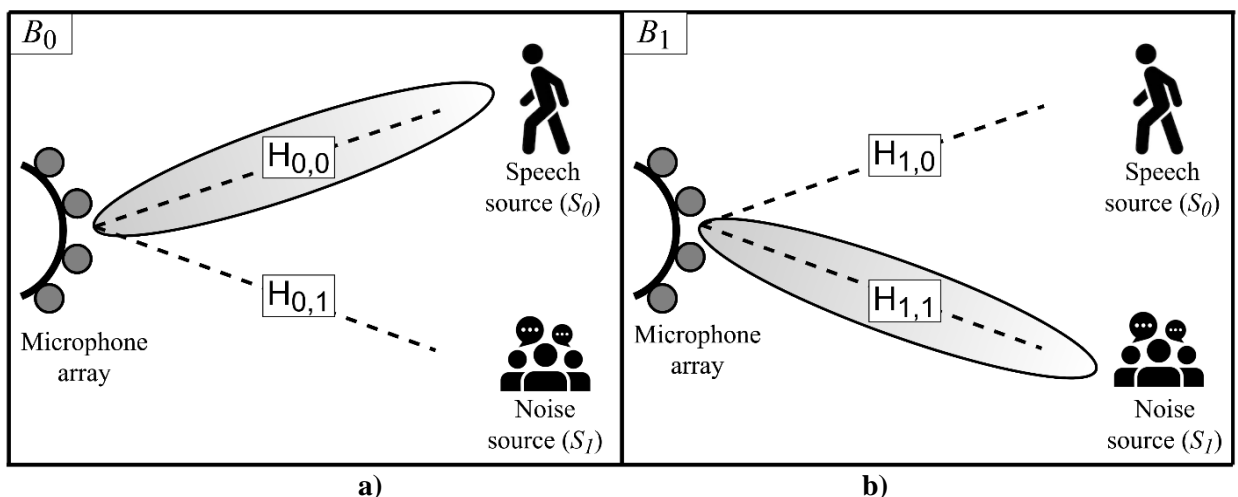

**Figure 2:** Beamforming for target speech source (a) and beamforming for target noise source (b).

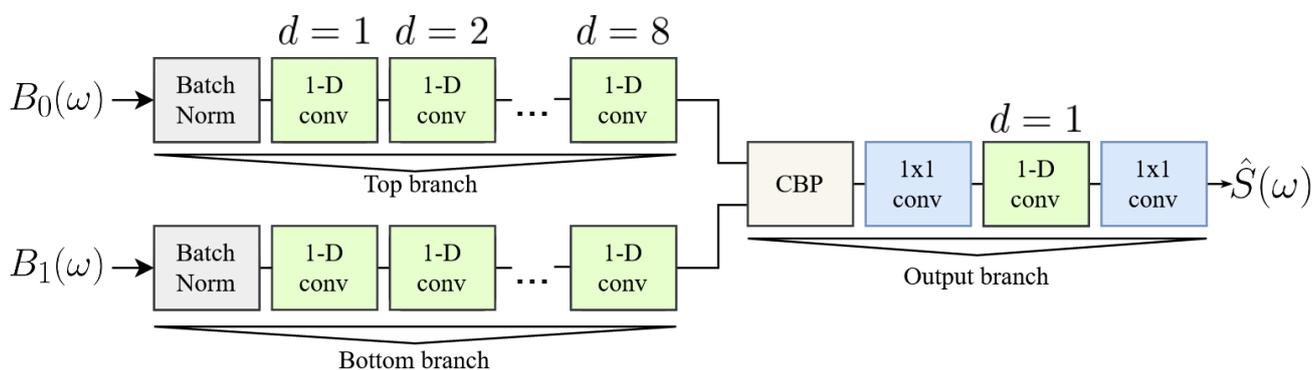

**Figure 3:** Our proposed deep-learning architecture for speech source separation.

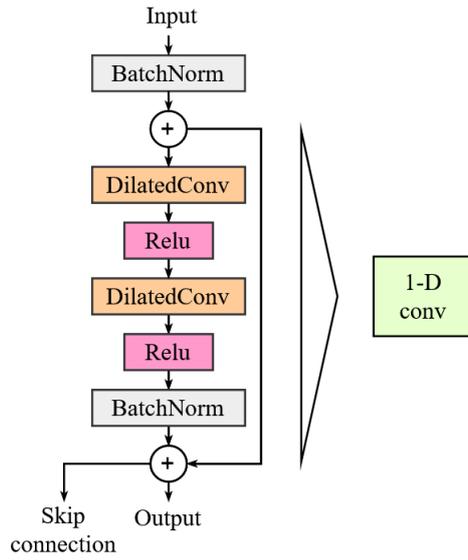

**Figure 4:** Detail of the one-dimensional convolution block in Fig. 3 of the proposed deep learning architecture.

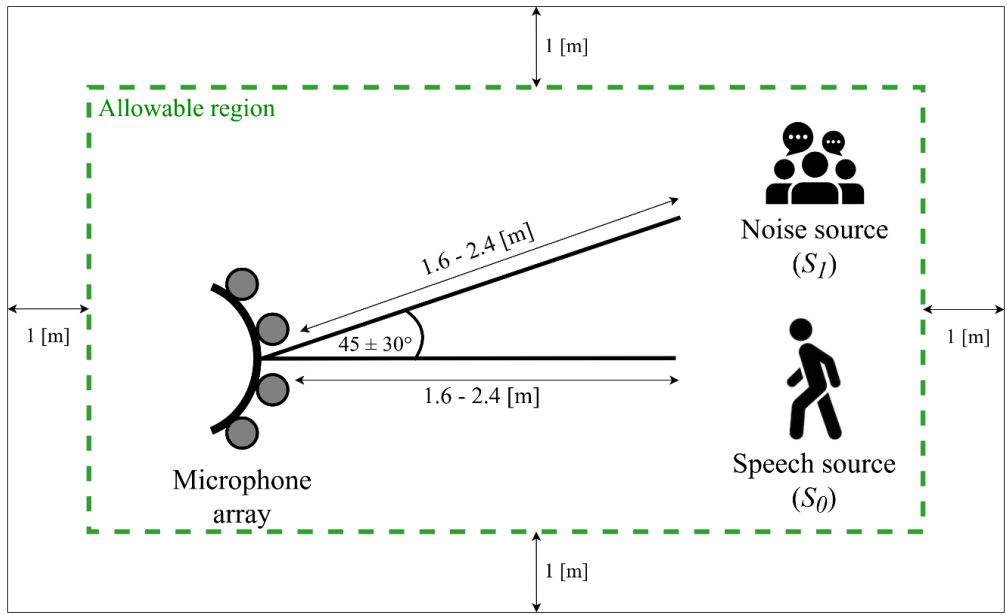

**Figure 5:** Scenario simulated with Pyroomacoustics to generate the RIRs employed in this research.

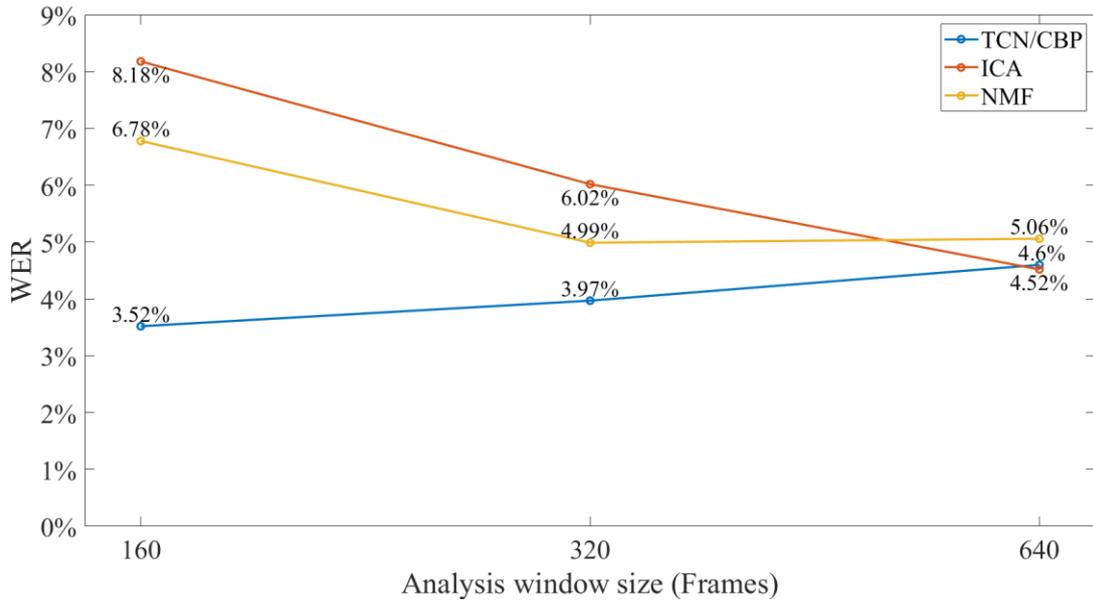

**Figure 6:** WER as a function of the analysis window size in number of frames. No reverberant condition was employed. The experiments were carried out with the clean trained ASR system.

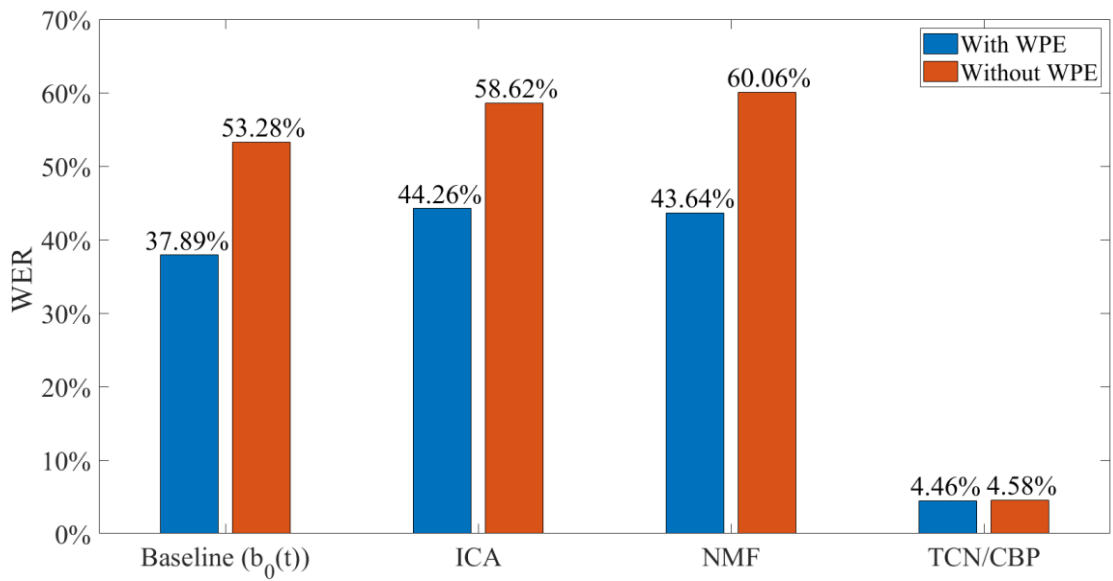

**Figure 7:** WER source separation experiments in matched reverberant environments, RIR-Matched. The same RIR quadruple was used to train and test the proposed TCN/CBP, i.e. RIR-Matched. TCN/CBP was applied every 160-frame analysis window. In contrast, ICA and NMF ran on the whole utterances. The experiments were carried out with the clean trained ASR system.

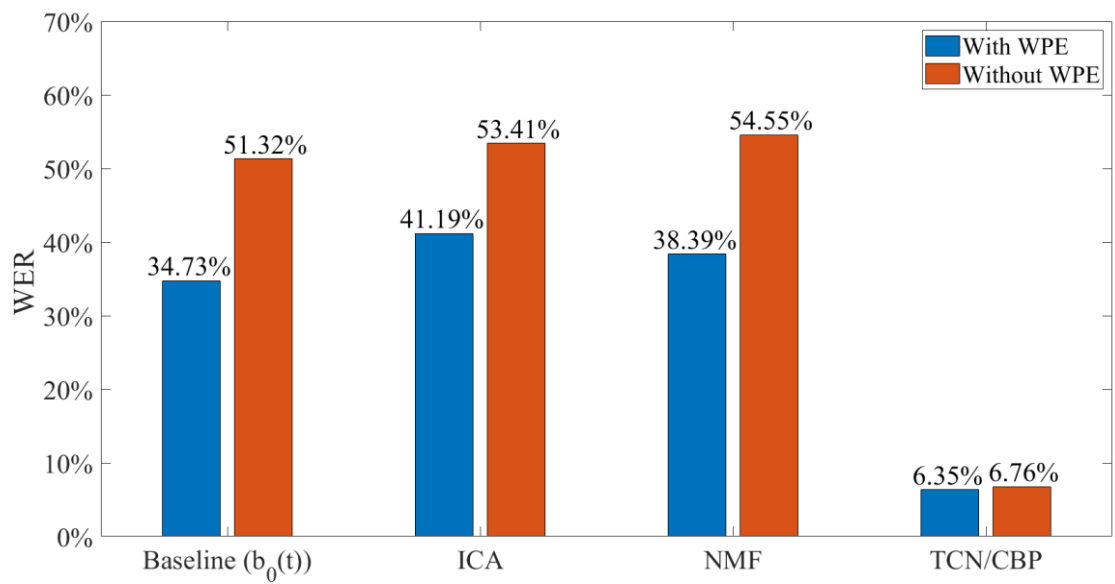

**Figure 8:** WER source separation experiments in reverberant environments. Multiple RIR quadruples were used to train and test the proposed TCN/CBP, i.e. RIR-Multicondition. TCN/CBP was applied every 160-frame analysis window. In contrast, ICA and NMF ran on the whole utterances. The experiments were carried out with the clean trained ASR system.

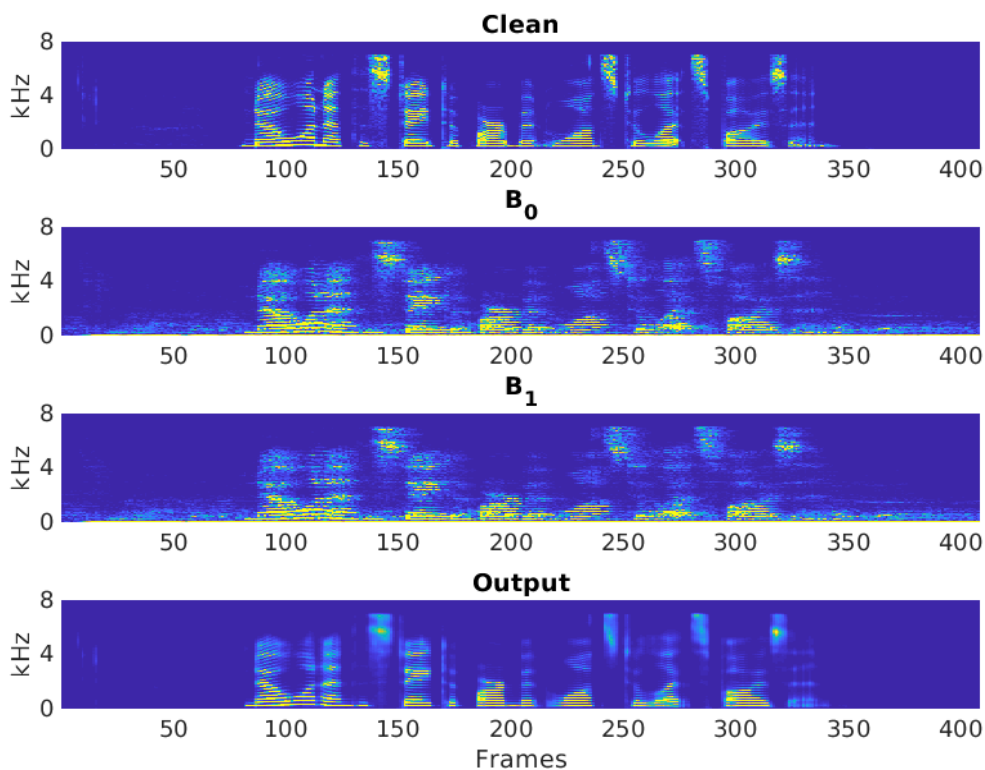

**Figure 9:** Spectrograms obtained with the target clean signal, beamforming signals $b_0(t)$ and $b_1(t)$, and the clean signal estimated with the TCN/CBP scheme and RIR-Multicondition.

**Table 1:** WER with source separation experiments without reverberation. TCN/CBP uses an analysis window equal to 160 frames, and ICA and NMF employed the whole utterance. The experiments were carried out with a clean trained ASR system.

| Method | WER |
|---|---|
| Baseline with $B_0(\omega)$. | 15.90% |
| ICA | 3.74% |
| NMF | 2.99% |
| TCN/CBP | 3.52% |

**Table 2:** Results with source separation in reverberant conditions as a function of the analysis window length. The experiments were carried out with a clean trained ASR system.

| | Analysis window size (frames) | | |
|---|---|---|---|
| **Testing condition** | **160** | **320** | **640** |
| RIR-Matched TCN/CBP | 4.58% | 4.60% | 4.90% |
| RIR-Multicondition TCN/CBP | 6.76% | 6.73% | 7.08% |
| RIR-Matched TCN/CBP + WPE | 4.46% | 4.57% | 4.74% |
| RIR-Multicondition TCN/CBP + WPE | 6.35% | 6.10% | 6.75% |
| Time-Varying-SNR TCN/CBP | 7.46% | 7.46% | 7.76% |

**Table 3:** Comparison of CBP with simple feature concatenation in Fig. 3. Also, the source separation scheme in Fig. 3 is compared with using only one beamforming signal, i.e. $B_0(\omega)$ or $B_1(\omega)$. The experiments were carried out with the RIR-Multicondition sets and the clean trained ASR system.

| | Analysis window size (frames) | | |
|---|---|---|---|
| **Source separation condition** | **160** | **320** | **640** |
| TCN/CBP | 6.76% | 6.73% | 7.08% |
| TCN/Concat | 6.87% | 6.61% | 7.10% |
| TCN with $B_0(\omega)$ only | 7.34% | 6.76% | 7.68% |
| TCN with $B_1(\omega)$ only | 9.25% | 8.94% | 9.26% |

**Table 4:** Comparison of CBP with simple feature concatenation in Fig. 3. Also, the source separation scheme in Fig. 3 is compared with using only one beamforming signal, i.e. $B_0(\omega)$ or $B_1(\omega)$. The experiments were carried out with the Time-Varying-SNR set and the clean trained ASR system.

| Testing condition | Analysis window size (frames) | | |
|---|---|---|---|
| | **160** | **320** | **640** |
| TCN/CBP | 7,46% | 7,46% | 7,76% |
| TCN/Concat | 7,53% | 7,35% | 7,44% |
| TCN with $B_0(\omega)$ only | 7,95% | 7,42% | 8,36% |
| TCN with $B_1(\omega)$ only | 10,16% | 9,61% | 10,18% |

**Table 5:** Comparison of WERs achieved with clean and multi-condition trained ASR systems. The following testing conditions were evaluated: $b_0(t)$ and RIR-Multicondition; and the estimation of the clean signal with TCN/CBP, $\hat{s}(t)$, also with RIR-Multicondition.

| Testing condition | Clean trained ASR | Multi-condition trained ASR |
|---|---|---|
| $b_0(t)$, RIR-Multicondition | 51.32% | **7.75%** |
| $\hat{s}(t)$, TCN/CBP RIR-Multi-condition | **6.76%** | 20.29% |

**Table 6:** Comparison of the proposed TCN/CBP with ICA and NMF as evaluated here.

| | **ICA** | **NMF** | **TCN/CBP** |
|---|---|---|---|
| **Training time** | NA | NA | 1 hr. |
| **Testing time (seconds per utterance).** | 0.07s | 55 s | 0.03s |
| **Dependence on the analysis window size** | High | High | Low |
| **Robust to reverberation** | No | No | Yes |